\documentstyle[12pt]{article}
\begin{document}

\title{Diffusion process in a deterministically forced flow}
 
\author{Piotr Garbaczewski\thanks{Presented by P. Garbaczewski at the
X Symposium on Statistical Physics, Zakopane, September 1-10, 1997}\\
Institute of Theoretical Physics, University of Wroc{\l}aw,
pl. M. Borna 9, \\
PL-50 204 Wroc{\l}aw, Poland}

\maketitle
\hspace*{1cm}

\begin{abstract}
We analyze  circumstances  under which  the  microscopic dynamics
 of particles which are driven by  a forced, gradient-type flow
can be consistently interpreted as a Markovian diffusion process.
 Special attention is paid to discriminating between forces that
 are presumed to act selectively upon diffusing particles,  while
 leaving the random medium statistically at rest
 (Smoluchowski diffusion processes), and those perturbing the
 random medium  itself and thus creating the nontrivial flows.
 We focus on the deterministic "stirring" scenarios.
\end{abstract}
\vskip1.0cm

To  analyze  random
perturbations that are either superimposed upon or are intrinsic to a
driving deterministic  motion, quite typically
a configuration space equation
$${\dot{\vec{x}}=\vec{v}(\vec{x},t)}\eqno (1)$$
 is invoked,  which  is next replaced by
a formal infinitesimal representation  of   an It\^{o} diffusion 
process
$${d\vec{X}(t)= \vec{b}(\vec{X}(t),t)dt + \sqrt{2D} d\vec{W}(t)
\enspace .}\eqno (2)$$
Here,  $\vec{W}(t)$  stands for  the normalised  Wiener noise, and
$D$  for a diffusion constant.  

The dynamical meaning of $\vec{b}(\vec{x},t)$, and thus reasons for
 making  a substantial difference between the forward drift  of the
process and the driving velocity field (1),
 relies on a  specific  diffusion input and  its possible phase-space
 (e.g. Langevin or that coming from deterministic dynamical systems)
  implementation,  that  entail  a detailed  functional
relationship of $\vec{v}(\vec{x},t)$ and $\vec{b}(\vec{x},t)$,
and justify  such notions like:  diffusion in an external force field,
  diffusion under various strains, diffusion along,
     against or across the driving deterministic flow, \cite{horst}.
We shall \it not \rm touch upon an important issue of diffusion under
shear, \cite{horst1}, when nontrivial vortices may arise, by assuming
from the very beginning that  only the gradient velocity fields and
deterministic forces are  of interest for us in  the present paper.

The pertinent mathematical formalism  corroborates 
both the Brownian motion of a single particle in flows of various
origin  and  the diffusive
transport of neutrally buoyant components in flows of the
hydrodynamic type.
However, our  major issue is a probabilistic interpretation of various
linear and nonlinear partial differential equations of physical
relevance, hence with a slightly abstract flavour put against the
generally favoured practical reasoning.
  Expressing that in more physical terms,
 we address  an old-fashioned  problem of "how much nonlinear",
  "how much time-dependent", and generally--"how much arbitrary"
  can be the driving velocity field to yield a consistent stochastic
 diffusion process (or the Langevin-type dynamics). Another issue is to
 get hints about a possible  non-deterministic origin
 of such fields, \cite{vigier}.

Clearly, in random media that are 
statistically at rest, diffusion of single tracers or dispersion of 
pollutants are well described by the Fickian outcome of the molecular 
agitation, also in the presence of external force fields (then,
in terms of Smoluchowski diffusions).  On the other hand, it is
of fundamental importance to understand how
flows  in a random medium  (fluid, as example) affect dispersion.
Such velocity fields are normally postulated as a priori given
agents in  the formalism and their (molecular or else) origin is
disregarded. Moreover, usually the force exerted upon tracers
is viewed independently from
the forcing ("stirring") that might possibly perturb  the
random medium itself and create nontrivial (driving) flows.

Except for suitable continuity and growth restrictions, necessary to
guarantee  the existence  of the process $\vec{X}(t)$ governed by
 the It\^{o} stochastic
differential equation, the choice of the driving velocity field
$\vec{v}(\vec{x},t)$ and hence of the related drift
$\vec{b}(\vec{x},t)$ is normally  (in typical physical problems)
regarded   to be arbitrary (except for being "not too nonlinear", see
however at van Kampen's discussion of that issue in
Ref. \cite{horst}).

The  situation looks deceivingly simple, \cite{horst1},
if  we are (for example)
interested in a diffusion process  interpretation of passive tracers
dynamics in the  \it a priori \rm given
flow  whose velocity field is  a solution of the nonlinear
partial differential equation, be it Euler,
Navier-Stokes, Burgers or the like.
An implicit  assumption, that passively buoyant tracers in a fluid
have a
negligible effect  on the flow, looks  acceptable (basically,
in case when
the concentration of a passive component in a flow is small).
Then,  one is tempted   to view  directly
 the fluid velocity field $\vec{v}(\vec{x},t)$
as  the forward drift $\vec{b}(\vec{x},t)$ of the process,
with the contaminant
being diffusively dispersed along the streamlines.

Here, apparent problems arise:
irrespectively  of a specific physical context and the phenomenology
(like e.g. the Boltzmann equation with its, as  yet, not well
  understood Brownian motion approximation) standing behind the
  involved  partial
differential equations, some
stringent mathematical criterions must be met to justify the
diffusion process scenario, be it merely a crude approximation
of reality.

Namely, in  general,  the assumed nonlinear evolution rule for
$\vec{v}(\vec{x},t)$ must be checked against the dynamics
that is allowed to govern the space-time dependence of the
forward drift
field $\vec{b}(\vec{x},t)$ of the pertinent  process, \cite{nel},
which  is \it not \rm at all arbitrary.
The  latter is ruled by  standard consistency  conditions 
that are respected by any 
 Markovian diffusion process, and additionally by  the rules of the 
forward and backward It\^{o} calculus, \cite{horst,nel}, the
mathematical input that is frequently ignored in the physical literature.

Normally, the pragmatically oriented authors do not pay any attention
to such problems and feel free to use any (deterministically or not)
motivated velocity fields as forward drifts. In that case, serious
troubles  follow.

Indeed, the closely related  issue we have analyzed
before, \cite{burg}, where
as a by-product of the discussion, the  forced Burgers dynamics
$${\partial _t\vec{v}_B + (\vec{v}_B\cdot \vec{\nabla })\vec{v}_B=
D \triangle \vec{v}_B +
\vec{\nabla }\Omega }\eqno (3)$$
and  the  diffusion-convection equation
$${\partial _t c +
(\vec{v}_B\cdot \vec{\nabla })c = D\triangle c}\eqno (4)$$
(originally,  for  the concentration
$c(\vec{x},t)$ of a passive  component in a flow),
in case of gradient velocity fields, were found  to be
generic to a  Markovian diffusion process input  and as generically
incompatible with the standard continuity equation in the compressible
regime.
In that case, the dynamics of
concentration (in general this notion \it does not \rm coincide with
the probability density !) results from
 the  stochastic  process whose density $\rho (\vec{x},t)$
evolves according to  the standard Fokker-Planck equation
 $${\partial _t \rho =
 D\triangle  \rho - \vec{\nabla }\cdot (\vec{b}\rho )\enspace ,}
 \eqno (5)$$
 the   forward  drift solves an evolution equation:
$${\partial _t\vec{b} +(\vec{b}\cdot \vec{\nabla })\vec{b} =
- D\triangle \vec{b}
+ \vec{\nabla } \Omega \enspace ,}\eqno (6)$$
and there holds
 $${\vec{b} \doteq
 \vec{v}_B +  2D\vec{\nabla }ln \rho \enspace .}\eqno (7)$$

By combining  intuitions which underly the self-diffusion
description,
\cite{spohn},  with those appropriate for  probabilistic solutions
of the
so-called Schr\"{o}dinger boundary-data and next-interpolation
problem, \cite{burg,olk,zambr},
 the above argument can be generalized to  arbitrary
conservatively forced  diffusion processes, quite irrespectively of a
physical context in which their usage can be justified.

Namely, let us consider a  density
$\rho (\vec{x},t), t\geq 0$
of a stochastic diffusion process, solving   the Fokker-Planck
equation (5).
  For drifts that are gradient fields, the potential
$\Omega $ in Eqs. (3) and (6)  (\it whatever \rm its functional form is),
\it must \rm allow for  a  representation  formula, reminiscent of
the probabilistic Cameron-Martin-Girsanov transformation:
$${\Omega (\vec{x},t) = 2D[ \partial _t\Phi + {1\over 2}
({\vec{b}^2\over {2D}} + \vec{\nabla }\cdot \vec{b})]\enspace ,}
\eqno (8)$$
 where $\vec{b}(\vec{x},t) = 2D \vec{\nabla } \Phi (\vec{x},t)$.
The formula (8) is a trivial identity, if we take for granted that
all drifts are known from the beginning, like in case of typical
Smoluchowski diffusions.  Nonetheless, we always end up with
a concrete   space-time dependent function $\Omega (\vec{x},t)$
which enters the partial differential equation (6).
If we take Eq. (6) as a starting point with $\Omega $ a priori given,
its solutions
may be then sought for in turn (to yield the previous a priori
given drifts, if the procedure is consistent).

Also, the functional properties of $\Omega (\vec{x},t)$ are not
 an innocent feature of the formalism, since
for the existence of
the Markovian diffusion process with the forward drift
$\vec{b}(\vec{x},t)$, we must  resort to  potentials
$\Omega (\vec{x},t)$ that are  \it not  \rm completely  arbitrary
functions.
Technically, \cite{olk}, the minimal requirement is that the
admissible potential is a
 bounded from below continuous function.
 This restriction will have profound consequences for
 our further discussion of diffusion in a flow, although nothing
 serious happens if  $\Omega $ is bounded and, for example,
 is the periodic space-time function.

{\bf Remark 1:} If we set  $\rho = \rho _1 + \rho _2$, and demand that
$\rho _1 \neq \rho $  solves the Fokker-Planck equation with
the very
same drift  $\vec{b}(\vec{x},t)$ as $\rho $ does, then as a
necessary consequence of the general
formalism, \cite{burg,olk},  the concentration
$c(\vec{x},t)={{\rho _1(\vec{x},t)}\over {\rho (\vec{x},t)}}$
solves an associated diffusion-convection equation $\partial _t c +
(\vec{v}_B\cdot \vec{\nabla })c= D\triangle c$.
Here,  the flow velocity
$\vec{v}_B(\vec{x},t)$ coincides with the backward drift
 of the generic  diffusion process
with the   density $\rho (\vec{x},t)$.\\

We should clearly discriminate between forces whose effect is a
"stirring" of the random medium and those acting selectively on
diffusing
 particles, with a negligible effect on the medium itself.
 For example, the traditional Smoluchowski diffusion processes  in
  conservative force fields  are considered  in random media that are
  statistically at rest.  Following the standard (phase-space,
  Langevin) methodology, let  us
  set $\vec{b}(\vec{x})={1\over \beta }\vec{K}(\vec{x})$,
  where $\beta $
  is a (large) friction coefficient  and $\vec{K}$ represents an
   external Newtonian force per unit of mass ( e.g. an acceleration)
  that is of gradient from, $\vec{K}=-\vec{\nabla }U$.
  Then, the effective
  potential $\Omega $ reads:
  $${\Omega  = {\vec{K}^2\over {2\beta ^2}}+ {D\over \beta }
  \vec{\nabla }\cdot \vec{K}}\eqno (9)$$
and the only distinction  between the
  attractive or repulsive cases  can be read out from
     the term  $\vec{\nabla }\cdot \vec{K}$.
  For example, the harmonic attraction/repulsion
  $\vec{K}=\mp  \alpha \vec{x}, \, \alpha >0$ would give rise to
  a harmonic  repulsion, if interpreted in terms of
   $\vec{\nabla }\Omega$, in view of
  $\Omega = {\alpha ^2\over
  {2\beta ^2}}\vec{x}^2 \mp  3D{\alpha \over \beta}$.
The situation would not  change under  the
incompressibility condition (cf. also the probabilistic approaches to
the Euler, Navier-Stokes and Boltzmann equations, \cite{marra}).

Notice that by formally changing a sign of $\Omega $ we would arrive at
the attractive variant of the problem, which  is however \it
incompatible  \rm
with the diffusion process scenario in view of the unboundedness of
$- \Omega $ from below.

We have thus arrived at the major point of our discussion:
 a priori, there is no way to incorporate the attractive forces
which affect (drive) the flow and  nonetheless generate a consistent
diffusion-in-a-flow  transport.

Clerly, there is no reason to  exclude the attractive variants of
the potential
$\Omega $ from considerations, since the deterministic motion is
consistent with them.

Concluding,  if the diffusion is to be involved
 we  need to save the situation somehow, and this can be made
 only by incorporating the
hitherto not considered "pressure" term effects.That is
suggested by the general form of the compressible Euler
($\vec{F}=-\vec{\nabla }V$ stands for external volume forces and
 $\rho $ for the fluid density that \it itself \rm undergoes
 a stochastic diffusion process):
$${\partial _t\vec{v}_E + (\vec{v}_E\cdot \vec{\nabla })\vec{v}_E =
\vec{F} - {1\over {\rho }}\vec{\nabla }P}\eqno (10)$$
or  the incompressible, \cite{marra}, Navier-Stokes equation:
$${\partial _t\vec{v}_{NS} + (\vec{v}_{NS}\cdot \vec{\nabla })
\vec{v}_{NS} =  {\nu \over {\rho }}\triangle \vec{v}_{NS} + \vec{F} -
{1\over{\rho }}\vec{\nabla }P\enspace ,}\eqno (11)$$
both to be compared with the equations (1) and (4), that set
 dynamical constraints  for respectively backward and forward drifts
 of a Markovian diffusion process ?

Let us stress again  that the acceleration term $\vec{F}$ in
equations (10) and (11) normally is regarded   as arbitrary,
while the corresponding term
$\vec{\nabla }\Omega $   in (3), (6)  and (8)  involves a bounded from
below function.

Since, in case of gradient velocity fields, the dissipation term
in the incompressible  Navier-Stokes equation (11) identically
vanishes,
we should concentrate on analyzing the possible
"forward drift of the Markovian process"  meaning of the Euler
flow with the velocity field   $\vec{v}_E$,  (10).

At this point it is useful, at least on the formal grouds, to invoke
the standard phase-space argument that is valid for a Markovian
diffusion process
taking place in a given flow  $\vec{v}(\vec{x},t)$ with as yet
unspecified dynamics nor physical origin.
We account for an explicit force exerted upon
diffusing particles, while not necessarily directly
affecting the driving flow itself.
Namely, \cite{horst1,nel}, let us  set for infinitesimal increments of
phase space random variables:
$$d\vec{X}(t)= \vec{V}(t) dt $$
$${d\vec{V}(t)= \beta [\vec{v}(\vec{x},t) - \vec{V}(t)] dt +
\vec{K}(\vec{x})dt  + \beta \sqrt{2D} d\vec{W}(t)\enspace .}
\eqno (12)$$

Following the leading idea of the Smoluchowski approximation, we
assume
 that $\beta $ is large, and consider the process for  times
 significantly exceeding $\beta ^{-1}$. Then, an appropriate
  choice of
 the velocity field $\vec{v}(\vec{x},t)$
 (boundedness and  growth restrictions  are  involved) may
 in principle guarantee, \cite{nel},  the convergence of the
 spatial part
 $\vec{X}(t)$ of the process (12) to  the It\^{o} diffusion process
 with infinitesimal increments (where the  force $\vec{K}$ effects
 can be safely ignored if we are interested mostly in the driving
 motion):
 $${d\vec{X}(t) = \vec{v}(\vec{x},t)dt + \sqrt{2D}
 d\vec{W}(t)\enspace .}\eqno (13)$$

However, one cannot blindly insert in the place of the forward drift
$\vec{v}(\vec{x},t)$ any of the previously considered bulk
velocity fields,
without  going into  apparent contradictions.
Specifically, the equation (4) with
$\vec{v}(\vec{x},t)\leftrightarrow \vec{b}(\vec{x},t)$   must be
valid.

By resorting to velocity fields $\vec{v}(\vec{x},t)$  which obey
$\triangle \vec{v}(\vec{x},t)=0$, we may pass from (6) to  an equation
of the Euler form, (10), provided (8) holds true  and then
the right-hand-side
of (6) involves a bounded from below effective potential $\Omega $.

An additional requirement is that
$${\vec{F}-{1\over {\rho }}\vec{\nabla }P \doteq  \vec{\nabla }
\Omega \enspace .}\eqno (14)$$

Clearly, in case of a constant pressure we are left with the dynamical
constraint ($\vec{b}\leftrightarrow \vec{v}_E$):
$${\partial _t \vec{b} + (\vec{b}\cdot \vec{\nabla })\vec{b}
= \vec{F}=\vec{\nabla }\Omega } \eqno (15)$$
combining simultaneously the Eulerian fluid and the Markov diffusion
process inputs, \it if and only if \rm
$\vec{F}$ is repulsive, e.g. $- V(\vec{x},t)$ is  bounded from
below.
Quite analogously, by setting $\vec{F}=\vec{0}$, we would get
a constraint
on the admissible pressure term, in view of:
$${\partial _t\vec{b} + (\vec{b}\cdot \vec{\nabla })\vec{b} =
- {1\over \rho }\vec{\nabla }P = \vec{\nabla }\Omega \enspace .}
\eqno (16)$$

Both, in cases  (15), (16) the effective potential $\Omega $ must
respect the functional dependence (on a forward drift and its
 potential)  prescription (8). In addition, the Fokker-Planck
 equation (5) with the forward drift $\vec{v}_E(\vec{x},t) \doteq
 \vec{b}(\vec{x},t)$ must be valid
 for the density $\rho (\vec{x},t)$.

To our knowledge,  in the literature there is  only one known specific
class of Markovian diffusion
processes that would render the right-hand-side of Eq. (10)
repulsive but nevertheless account for the troublesome  Newtonian
accelerations, e.g. those of the from $- \vec{\nabla }V$, with $+V$
bounded from below.
Such processes  have forward
drifts that for each suitable,  bounded from below function
$V(\vec{x})$ solve the nonlinear  partial differential equation:
$${\partial _t\vec{b} + (\vec{b}\cdot \vec{\nabla })\vec{b} =
- D\triangle \vec{b} +
\vec{\nabla }(2Q - V)}\eqno (17)$$
with the compensating pressure term:
$${Q \doteq 2D^2 {\triangle \rho ^{1/2}\over \rho ^{1/2}} \doteq
{1\over 2}\vec{u}^2 + D\vec{\nabla }\cdot \vec{u}}\eqno (18)$$
$$\vec{u}(\vec{x},t)= D\vec{\nabla } ln\, \rho (\vec{x},t)$$
Their  discussion can be found in Refs.
\cite{nel,burg,olk,zambr}.

Clearly,  we have:
$${\vec{F}=-\vec{\nabla }V\, , \,
\vec{\nabla }2Q = - {1\over \rho }\vec{\nabla }P}\eqno (19)$$
where:
$${P(\vec{x},t)= - 2 D^2 \rho(\vec{x},t)\, \triangle \,
ln\, \rho (\vec{x},t)}\eqno (20)$$
Effectively, $P$ is here defined up to a
time-dependent constant.
Another admissible form of the pressure term reads (summation
convention is implicit):
$${{1\over \rho }\vec{\nabla }_k[\rho \,  (2D^2 \partial _j
\partial _k)
ln\, \rho ]  =
\vec{\nabla }_j (2Q)}\eqno (21)$$.

If we consider a subclass of processes for which the dissipation term
identically vanishes ( a number of examples can be found in Refs.
\cite{olk}):
$${\triangle \vec{b}(\vec{x},t)=0}\eqno (22)$$
the equation (17) takes a conspicuous Euler form  (10), $\vec{v}_E
\leftrightarrow  \vec{b}$.

Let us notice that  (20), (21) provide for a  generalisation  of
the more
 familiar, thermodynamically motivated and suited for ideal gases and
 fluids,   equation of
 state $P \sim \rho $.  In case of density fields  for which
 $-\triangle ln\, \rho \sim const$, the standard relationship between
 the pressure and the density is reproduced.
 In case of density fields obeying
 $-\triangle ln\, \rho =0$, we are left with at most  purely time
 dependent  or a constant pressure. Pressure profiles may be highly
 complex for arbitrarily chosen initial density and/or the
 flow velocity fields.

To conclude the present discussion let us invoke Refs.
\cite{marra,spohn,olk}.
  The problem of a   diffusion process interpretation
of various partial differential equations  is known to  extend
beyond the original  parabolic equations setting, to general
nonlinear velocity field equations.
On the other hand, the nonlinear Markov jump processes associated with
the Boltzmann equation,
in the hydrodynamic limit, are believed  to imply either an ordinary
deterministic  dynamics with
 the velocity field solving the Euler equation, or  a diffusion
process
whose drift is a solution of the incompressible Navier-Stokes
equation
(in general, without our  $curl \vec{v}=0$ restriction),
\cite{spohn,marra}.
The case of \it arbitrary \rm external forcing has never  been
satisfactorily solved.

Our reasoning went otherwise.  We asked for the admissible space-time
dependence  of general velocity fields that are to play the r\^{o}le
of forward drifts  of Markovian diffusion processes, and at the
same time can be met in physically signicant contexts. Therefore
various forms of the Fokker-Planck equation  for tracers
driven by familiar compressible velocity fields  were  discussed.

Our finding is that
solutions of the compressible Euler equation  are appropriate for
the description of the general  non-deterministic
(e.g. random and Markovian)
dynamics running under the influence of  both  attractive and
repulsive stirring forces,
and refer  to a class of Markovian diffusion processes
orginally introduced by E. Nelson, \cite{nel,olk,vigier}.
That  involves only the gradient velocity fields
(a couple of issues concerning the $curl \, \vec{b} \neq 0 $
velocity
fields and their nonconservative forcing have  been raised
in Refs. \cite{burg}).

{\bf Remark 2:} Let us stress that a standard justification
of the hydrodynamic limit for a tracer   particle
invokes a Brownian particle in an equilibrium  fluid.
An issue of how
much the tracer  particle disturbs the fluid (random medium)
locally and how far away  from the tracer particle  the thermal
equlibrium
conditions regain their validity, \cite{spohn},  normally is
disregarded.
Moreover, in the standard derivation of local conservation laws
from the Boltzmann equation, the forcing term on the
right-hand-side of the Euler or Navier-Stokes equation   up to
scalings does
coincide with the force acting on each single particle comprising
the system.     Thus, in this framework, there is no room for any
discrimination
between forces acting upon tagged particles and those perturbing the
spatial flows (once on the level of local averages).

Quite on the contrary,
the force term in the Kramers equation and this appearing in the
related local conservation law for the forward drift or for the
current velocity of the diffusion process are known
\it not \rm to coincide in general.  Typically, the action of an
external force  is confined  to diffusing (tagged) particles  with
\it no \rm global or local effect on the surrounding random medium,
cf. standard
derivations of the Smoluchowski equation.
This feature underlies problems with the diffusion process
interpretation of general partial differential equations governing
physically relevant velocity fields.
Specifically, any  external intervention (forcing)  upon
a stochastically evolving (in the diffusion process approximation)
system  gives rise to a perturbation of local flows, which seldom
can be
analyzed as forcing of any definite type on the molecular level.
The Smoluchowski theory is here a notable exception, but
there one has  no room for genuine flows and velocity field profiles
which are generated in the random medium. \\

{\bf Remark 3:}  It seems worthwhile to mention a close connection
of the considered framework with  the general issue of executing small
random perturbations on the level of the classical Hamilton-Jacobi
dynamics, \cite{zambr}, with the related issue of an optimal control of
stochastic processes and with that of the so-called "viscosity
solutions" of the Hamilton-Jacobi equation, \cite{fleming}.
In fact, our (Feynman-Kac, see
\cite{olk}) potentials (8), (9) were introduced on the basis of
probabilistic  arguments via the Girsanov or Cameron-Martin theorems
about transformations of drifts of the diffusion process. However,
an  implicit assumption that drifts are defined in terms of
gradients of suitable logarithmic functions: $\vec{b}=2D
\vec{\nabla } log \theta $ and $\vec{v}_B\equiv \vec{b}_*=
-2D\vec{\nabla } log {\theta }_*$ (here, we employ  the notation of our
previous publications, \cite{olk}, where ${\theta }_*$ is
a bounded solution of the forward generalized diffusion equation,
while $\theta $ that of its time adjoint) implies that
the compatibility condition (8)  can be rewritten in two equivalent
forms, both involving  the  modified Hamilton-Jacobi equations.
Namely, let us
set $2D\, log \theta =\Phi $ and  $-2D\, log \theta _*=\Phi _*$.
Then, we have:
$\Omega = \partial _t \Phi + {1\over 2}|\vec{\nabla } \Phi |^2 +
D\triangle \Phi $ and at the same time $\Omega = \partial _t \Phi _*
+ {1\over 2} |\vec{\nabla } \Phi _*|^2 - D\triangle \Phi _*$.
The latter one is identified as the so-called Hamilton-Jacobi-Bellmann
programming equation in the optimal control of  stochastic diffusion
processes, \cite{fleming,zambr},
and via the Hopf-Cole logarithmic transformation (take the gradient) is
linked to the Burgers equation (3). An issue of viscosity solutions of
the  standard Hamilton-Jacobi equation has been extensively studied
in the literature as the
$D\downarrow  0$ limit of solutions of the modified (e,.g. Bellmann)
equation.  It is thus clear, on the basis of our previous discussion,
  that an apparent  obstacle is hidden
in the assumption that a  diffusion  process
is involved. Then, suitable restrictions upon $\Omega $ must be
respected, and the attractive versus repulsive potential problem appears.

\end{document}